\documentclass[aps,prd,letterpaper]{revtex4}

\usepackage{latexsym}
\usepackage{epsfig}
\usepackage{scrextend}
 \usepackage[french]{babel}

\usepackage{url}
\usepackage{hyperref}
\usepackage{latexsym}
\usepackage{epsfig}
\usepackage{scrextend}
\usepackage{lipsum}

\usepackage{natbib}

\usepackage{tikz}
\usetikzlibrary{positioning,arrows,decorations.pathmorphing,arrows.meta}

\usepackage{xcolor}
\usepackage{amssymb}
\usepackage{amsmath}
\usepackage{graphicx}

 \definecolor{blue}{RGB}{7,80,201}
 \definecolor{red}{RGB}{200,20,1}

\def\be{\begin{equation}}
\def\ee{\end{equation}}
\def\bea{\begin{eqnarray}}
\def\eea{\end{eqnarray}}
\def\ba{\begin{array}} 
\def\ea{\end{array}}
\def\bc{\begin{center}}
\def\ec{\end{center}}

\def\ghost#1{}

\def\simge{\mathrel{%
   \rlap{\raise 0.511ex \hbox{$>$}}{\lower 0.511ex \hbox{$\sim$}}}}
\def\simle{\mathrel{
   \rlap{\raise 0.511ex \hbox{$<$}}{\lower 0.511ex \hbox{$\sim$}}}}

\def\dis{\displaystyle}

\newcommand{\footremember}[2]{%
    \footnote{#2}
    \newcounter{#1}
    \setcounter{#1}{\value{footnote}}%
}

\begin{document}

\title{\boldmath  {\em MICROSCOPE} \,limits on the strength of a new force
\medskip \\  with comparisons to gravity and electromagnetism
\vspace{8mm}\\ 
}

\author{{\sc P}{ierre} {\sc Fayet}
\vspace{6mm} \\ \small }

\affiliation{Laboratoire de physique de l'\'Ecole normale sup\'erieure\vspace{.5mm}\\ \hspace*{4mm}24 rue Lhomond, 75231 Paris cedex 05, France 
\hspace{-1mm}\footremember{a}{\vspace*{0mm}   \em LPENS, \'Ecole normale sup\'erieure, Universit\' e PSL; CNRS UMR  8023;  
\vspace{.8mm}Sorbonne Universit\'e;    UPD-USPC
}
\vspace{1.5mm}\\
\hbox{and \,Centre de physique th\'eorique, \'Ecole polytechnique, 91128 Palaiseau cedex, France}
\vspace{2mm}\
\vspace{0mm}}

%%\date{January 23, 2019}  

\begin{abstract}

\textwidth 10cm

{\vspace{2mm}
\hspace*{2mm}Extremely weak new forces 
\vspace{.2mm} could lead to apparent violations of the Equivalence Principle.  The {\it MICROSCOPE\,}  experiment implies that the relative strength of a new long-range force, compared with gravity,
is constrained to $|\bar\alpha_g| <   \,3.2\times 10^{-11},\ 2.3 \times 10^{-13},\ 2.2 \times 10^{-13},\ 6.7\times 10^{-13}$ and $1.5\times 10^{-12}$ at $2\sigma$, for a coupling to $B,\ L,\ B\!-\!L,\ B\!+\!L$ or $3B\!+\!L$;
or, for a coupling  to isospin, $|\alpha_g| < \,8.4\times 10^{-12}$.
\,This is a gain in sensitivity $\simeq 3$  for a coupling to $B$,  to $\approx$ 15 in the other cases, including $\,\!B\!-\!L$ as suggested by grand unification.
\vspace{1.2mm}\\
\hspace*{3mm} 
This requires paying attention to the definition of  $\,\bar \alpha_g$.
A force coupled to  $L$ (or $B\!-\!L$) would act effectively on protons (or neutrons) only, its relative intensity being reduced from $\alpha_g$
to about $\,\bar\alpha_g=\alpha_g/4$  for an average nucleon.
A force coupled to $B+L=2Z+N$ would act twice as much on $p$ as on $n$, getting enhanced  from $\alpha_g$ for neutrons to about $\,\bar\alpha_g= \frac{9}{4}\,\alpha_g$ for an average nucleon.  
It is thus convenient to view such forces as acting on
$\bar Q =B,\ 2L,\ 2\,(B\!-\!L), \ 2\,(B\!+\!L)/3$ or $2\,(3B+L)/7$
(normalized to $2$  for  $p\!+\!e\!+\!n$),
leading to $\,\bar\alpha_g=\alpha_g  \times  (1,\,1/4,\,1/4,\,9/4$ or $49/4$). 
The sensitivity for a coupling to $L$ or $B\!-\!L$\, is better than for $B$  by {\em two orders of magnitude\,}
(as $\, \Delta (2L/A_r) \simeq $ $144\  \Delta (B/A_r)$ for Ti-Pt);
and about 3 or 7 times better than for $B\!+\!L$ or $\,3B\!+\!L$.
\vspace{1.2mm}\\
\hspace*{3mm} 
A coupling to $(\epsilon_BB+\epsilon_{Q_{\rm el}}Q_{\rm el})\,e$ should verify 
$|\epsilon_B| < 5\times 10^{-24}$;  similarly $|\epsilon_L|$ or $|\epsilon_{B\!-\!L}| $ $< .9\times10^{-24}$, $\ |\epsilon_{B\!+\!L}|< .5 \times 10^{-24},\  |\epsilon_{3B+L}|< .32\times 10^{-24}$ and $\,|\epsilon_{B-2L}| < \,2.6\times 10^{-24}$, \,implying a new
\linebreak
 inter\-action weaker than electromagnetism 
by more than $10^{46}$ to $10^{48}$. The resulting hierarchy between couplings, typically by $\simge 10^{24}$, may be related within supersymmetry with a large hierarchy in energy scales by  $\simge 10^{12}$. This  points to a $\,\sqrt \xi \,\approx 10^{16}\!$ GeV scale, asso\-ciated with a huge vacuum energy density that may be responsible for the inflation of the early Uni\-verse.
\vspace{5mm}
\ \vspace{-2mm} \\
\hspace*{2mm}
\vspace{-2.5mm}\\
\hbox{\hspace{110mm}LPTENS/18/12}
\vspace{-0.5mm}\\
}
\end{abstract}

\maketitle

\section{Introduction}

There are four known types of fundamental interactions in nature.
Strong, weak and electromagnetic interactions are mediated by the $SU(3)\times SU(2)\times U(1)$
gauge bosons of the Standard Model  \cite{sm}. Gravitation is well described, at the classical level, by General Relativity, and gravitational waves from merging black holes or neutron stars have been observed recently~\cite{gw}.
Still many questions remain unanswered. Why four types of interactions, with these symmetries, and do others exist\,?
How can one get a consistent theory of quantum gravity\,? 
What is dark matter made of, and how can dark energy be interpreted\,? What is 
responsible for the very fast inflation of the early Universe ? Why is gravity so weak at ordinary energies, as compared to the other interactions\,? Can interactions be unified, and at which energies\,?

\vspace{2mm}
Most attempts at a better understanding of these questions
involve new symmetries, new particles and interactions, and new energy scales.
One of the simplest possibilities involves an extra $U(1)$  within a
$SU(3)\times SU(2) \times U(1)\, \times $ extra-$U(1)$ gauge group, or $SU(5)\, \times$ extra-$U(1)$ for a grand-unified theory, as  may be present in extensions of the standard model, supersymmetric or not.
The resulting gauge boson may be very heavy, $\simge$ a few TeV.
In a  less conventional situation, however, both the extra-$U(1)$ gauge coupling $\,g"\!$ and the mass of the corresponding gauge boson $U$ may be very small or extremely small \cite{plb80}, even down to $m_U=0$,
which could lead to an extremely weak new long-range force \cite{plb86,plb89}.
How can such a force fit within grand unification, what could its possible intensity be as compared with gravity and electromagnetism, and could it be related with a huge energy density at the origin of inflation\,? These are the questions we would like to discuss. 
\vspace{1.6mm}

New long-range forces adding their effects to those of gravity
are in general expected  to lead to apparent violations of the Equivalence Principle, at the basis of General Relativity. According to this principle, test bodies of different compositions should undergo the same free-fall accelerations, as investigated long ago by E\"otv\"os and his collaborators \cite{eot}. 
Such new forces, if long ranged, must be much weaker than gravitation, or they would have been discovered already \cite{adel1,adel2,adel3,adel4,fisch,will,safro}.
We shall discuss here the constraints following from the first results of the {\it MICROSCOPE\,} experiment testing the Equivalence Principle in space \cite{micro}, 
considering the exchanges of a spin-1 or spin-0 mediator, extending earlier results \cite{fayetmicro} and
 showing how the exact limits, as compared to gravitation, or electromagnetism, depend significantly on the coupling involved.  
For a force mediated by a very light or massless \hbox{spin-1} boson $U$, generalized version of a ``dark photon'', the coupling is expected 
from gauge symmetry (spontaneously broken or not) to be a linear combination of $B,\,L$ and $Q_{\rm el}$, expressed as $(\epsilon_B B+\epsilon_L L+\epsilon_{Q_{\rm el}}\,Q_{\rm el})\,e$. \,It is thus effectively given by $(\epsilon_B B+\epsilon_L L)\,e$,
and more specifically  $\epsilon_{B-L}\,(B-L)\,e\,$ in a grand-unified theory \cite{plb86,plb89,epjc}. The new force should then act proportionally to 
$B-L\,- .61~Q_{\rm el}$, in agreement with a high-energy $SU(4)$ {\it electrostrong symmetry}\, relating the photon with the eight gluons.

\vspace{1.6mm}
The intensity of the new force as compared to gravitation is better expressed in terms of a parameter $\bar\alpha_g$ averaging over protons and neutrons, rather than by the usual $\alpha_g$.
Indeed, a force effectively coupled to $L$ acts on protons (with accompanying electrons) but not on neutrons, leading to a relative strength parameter $\bar\alpha_g=\alpha_g/4$\,;\, and similarly for a force coupled to $B-L$, acting effectively on neutrons. 
It is thus convenient to reconsider such forces acting on $B,\ L,\,B\!-\!L,\,B\!+\!L$ or $3B\!+\!L$ as acting on the renormalized charges $\bar Q =B,\ 2L,\ 2\,(B\!-\!L), \ 2\,(B\!+\!L)/3$ or $2\,(3B+L)/7$ \,(all normalized in the same way to $\bar Q=2$  for  $p\!+\!e\!+\!n$),
with a relative strength compared to gravity 
$\,\bar\alpha_g=\alpha_g ,\ \alpha_g/4,\  \alpha_g/4,\ 9\, \alpha_g/4$ or $49\, \alpha_g/4$, rather than just $\alpha_g$.

\vspace{1.5mm}

The relative difference in the accelerations of two test masses is expressed by the E\"otv\" os parameter $\delta$, proportional to $\epsilon_B\,\Delta (B/A_r)+\epsilon_L\, \Delta(L/A_r)$.
As  $|\Delta(2L/A_r)|$ between the two Ti and Pt test masses is significantly larger than $|\Delta (B/A_r)|$  the limits on $\bar \alpha_g$, for a long-range force coupled to $L$ or $B-L$, are {\it two orders of magnitude stronger\,} than for a coupling to $B$. 
We give the limits for other combinations such as $B+L$, or  $3B+L$, possibly suggested by a Pati-Salam \cite{ps} symmetry, and $B-2L=N-Z$ (for which $\bar\alpha_g$ is no longer relevant). We also discuss the improvement brought by {\it MICROSCOPE\,} \cite{micro} over the earlier results from the E\" ot-Wash experiment \cite{adel1,adel2,adel3,adel4}.

\vspace{1.6mm}

The new force should thus be smaller than gravitation by more than $10^{10} $ to $10^{12}$, 
and  smaller than electromagnetism by about $10^{46}$  to $10^{48}$  at least. But is there any reason to consider such incredibly small forces, and why should we care about them\,?  \,Supersymmetry may give us a hint, or even a possible answer,  by providing a connection between a very small coupling and a very large energy scale. The extreme smallness of the extra-$U(1)$ coupling $g"$ associated with the new force, smaller than the electromagnetic coupling $e$ by $\,\simge 10^{23}$ to $10^{24}$, may be related  to an extremely large value for the mass$^2$ coefficient  $\xi$, parametrizing the $\xi D$ term \cite{fi} for the extra $U(1)$ in the Lagrangian density.  This $\xi D$ term also generates spin-0 mass$^2$  terms $\propto \xi g"$, typically $\simge$ a few TeV$^2$. A very small  $g"\to 0$ thus corresponds through this seesawlike mechanism to a very large $\xi\to\infty$ \cite{ssm}. It  provides a very large energy scale $\sqrt \xi \propto g"^{-1/2}$, typically $\simge 10^{16}$ GeV,  associated with a huge energy density that may be responsible for the very rapid inflation of the early Universe.

\section{{\boldmath {\em MICROSCOPE}\, limit on a new force coupled to $B$: \ $|\, \bar \alpha_{\lowercase{g}}| < \,$ {\normalsize $3.2\times10^{-11}$}}  {\it\em  (2$\sigma$)}}

The {\it MICROSCOPE\,} experiment provides at present the most stringent test on the validity of the Equivalence Principle \cite{micro}.
It constrains the E\" otv\" os parameter $\delta$  measuring the relative difference in the free-fall accelerations  of two test masses of Ti and Pt alloys to 
\be
\label{micro}
\delta\,(\hbox{Ti-Pt})=\,(\,-.1\pm.9 \,\hbox{(stat)} \pm .9\,\hbox{(syst)}\,)\times 10^{-14}\,.
\ee
%%The 1$\sigma$  statistical uncertainty is  $\,.9 \times 10^{-14}$ and the upper bound on the systematic uncertainty, dominated by thermal effects,  $.9 \times 10^{-14}$, the 1$\sigma$ systematic uncertainty being noticeably lower \cite{micro}.
This implies   that $ |\delta |$ should be smaller than about  $\,2.5 \times 10^{-14}$  \,at $2\sigma$  (and $ \,1.3\times  10^{-14}$ at $1\sigma$).

\vspace{2mm}
We denote for simplicity by Ti and Pt  the titanium and platinum alloys used for the 
 test masses (cf. Table \ref{table:grandeltamic}).
One usually intends to infer from there an upper limit on the relative strength of a new long-range force as compared to gravity, defining a parameter $\alpha_g$ associated with a modified Newton potential, often expressed as
%%\vspace{-4mm}
\be
\label{defalpha}
V(r)\, =\, -\ \frac{G_N\,m_a m_i}{r}\ (1+\alpha_g \,e^{-r/\lambda})\ .
\ee
$\lambda=\hbar/(mc)$ is the range of the new force, $m$ being the mass of its mediator, taken to be extremely small, or even 0.
But eq.\,(\ref{defalpha}) would require the new force to act exactly proportionally to mass, which is both unlikely (leaving aside the special  case of a spin-0 coupling to $T^{\mu}_{\ \mu}$) and not suitable when dealing with Equivalence Principle tests.

\vspace{2mm}

Defining an appropriate $\alpha_g$ requires some hypothesis on which quantity $Q$ the new force is supposed to act.
For a force acting proportionally to baryon number $B$ with a coupling constant $\,\epsilon_Be$,  \,the interaction potential reads

\vspace{-8mm}
\be
V_B(r)\,=\,\pm \ \epsilon_B^2 \ \frac{e^2}{4\pi\epsilon_\circ}\ B_aB_i \ \frac{e^{-r/\lambda}}{r}\ .
\ee
The upper and lower signs correspond to a spin-1 or spin-0 mediator, respectively.
The interacting masses may be expressed in atomic mass units  (u) as $m=A_r\,{\rm u} $, $A_r$ being the relative atomic mass of the element or macroscopic body  considered (normalized to $A_r=12$ for a $^{12}$C  atom).
We can write the modified Newton potential as

\vspace{-4mm}

\be 
\label{defalphab}
V(r) = \,-\,\frac{G_N\,m_a m_i}{r}\  \left[\,1+\alpha_{gB} \left(\frac{B}{A_r}\right)_{\!a}\,\left(\frac{B}{A_r}\,\right)_{\!i}\,e^{-r/\lambda}\,\right]\ ,
\ee
with \cite{fayetmicro}

\vspace{-8mm}
\be
\label{aleps}
\alpha_{gB}\,=\, \mp\ \frac{\alpha}{G_N\,{\rm u}^2}\  \epsilon_B^2\  
=\,\mp\ \alpha\ \left(\frac{m_{\hbox{\footnotesize Planck}}}{m_p} \right)^{\!2}\,\left(\frac{m_p}{\rm u}\right)^{\!2}\,\epsilon_B^2\,
\simeq\ \mp \ 1.2536\times 10^{36}\ \epsilon_B^2\,.
\ee

\vspace{2mm}

\begin{table}[t]
 \caption{\ The $Q/A_r$ ratios for the Ti  (TA6V$\!$: 90\% Ti, 6\% Al, 4\% V) and Pt (90\% Pt, 10\% Rh) {\it MICROSCOPE\,} test masses \cite{fayetmicro}, for forces acting proportionally to $Q=B, \,L,\,B\!-\!L,\,B\!+\!L, \,3B\!+\!L$ or $B\!-2L$\,.
 \vspace{2mm}\\ 
 \label{table:grandeltamic}}
  \begin{tabular}{c}
$\ba{|c||r|r||r|}
\hline 
&&& \\ [-2.5mm]
Q/A_r
& \hbox{Ti}_{\rm \,alloy}\ \ &\hbox{Pt}_{\rm \,alloy}\ \ &\ \  \Delta(Q/A_r)_{\hbox{\footnotesize \,Ti-Pt}} \ \ 
\\ [2mm]  \hline
&&& \\ [-2.5mm]
B/A_r
&\ \  1.00105 \ \   & \ \ 1.00026 \ \ &.00079 \hspace{8mm}
\\ [.5mm]
L/A_r
&\  .46061\ \  & .40357\ \  &.05704 \hspace{8mm}
\\ [.5mm]
(B-L)/A_r
&\  .54043 \ \ &  .59668 \ \ & -\,.05625 \hspace{8mm}
\\ [.5mm]
(B+L)/{A_r}
&\  1.46166 \ \ &  1.40383\ \  & \,.05783 \hspace{8mm}
\\ [.5mm]
\ \ (3B+L)/A_r\ \ 
&\  3.46376 \ \ & 3.40435 \ \ & \, .05941 \hspace{8mm}
\\ [.5mm]
\ \ (B-2L)/A_r\ \ 
&\  .07982\ \  & .19311 \ \ & \,-\, .11329 \hspace{8mm}
\\ [2mm]
\hline
\ea $
\end{tabular}
\vspace{3mm}
\end{table}

As  $B/A_r$ is close to 1$,\,\alpha_{gB}$ provides a good measure of the intensity of the new force, if long ranged,  as compared to gravity.  With $\,\Delta(B/A_r)_{\hbox{\footnotesize Ti-Pt}} \,\simeq \,.00079 $ \cite{fayetmicro,berge} (and $B/A_r\simeq 1.0008$ for the Earth) we can express the E\"otv\"os parameter as \cite{fayetmicro}
\be
\label{deltazero}
\delta_{\hbox{\scriptsize Ti-Pt}}
 \,\simeq\, .\,00079\ (\times 1.0008)\  \,\alpha_{gB}\ \simeq \ \mp\ 10^{33}\ \epsilon_B^2 \,.
\ee
A long-range force coupled to  $B$ must then verify
\be
\label{limalphab}
\framebox [10.2cm]{\rule[-.45cm]{0cm}{1.1cm} $ \dis
\ \left|\,\frac{\hbox{new force}}{\hbox{gravity}}\,\right|\ \simeq\ |\alpha_{gB}|\ < \ 
\hbox{$\dis \frac{2.5\times 10^{-14}}{.00079\ }$}
\ \simeq \ 3.2\times 10^{-11}\ \ \ (2\sigma)\,
$}
\ee
(or $\,<\,1.6\times 10^{-11}$ at 1$\sigma$).
\,And, in a sense specified later,
\be
\label{limalphab2}
\ \left|\ \frac{\hbox{new force}}{\hbox{electromagnetism}}\ \right|\ \,\simeq\ \epsilon_B^2\ < \ 
(2.5\times 10^{-14}) \times 10^{-33}
\ = \  2.5\times 10^{-47}\ \ \ (2\sigma)\,,
\ee
i.e. $|\,\epsilon_B|\,<\,5\times 10^{-24}$ \ (or $3.6\times 10^{-24}$ at 1$\sigma$) \cite{fayetmicro}.

\section{Comparison with the E\" ot-Wash experiment\hspace{-.5mm}}

 \begin{table}[t]
 \caption{\ $\Delta (Q/A_r)$ for the  Be-Ti and Be-Al \,E\"ot-Wash test masses, derived from \cite{adel4}. 
 We also give $\Delta(Q/A_r)_{\hbox{\scriptsize \,Ti-Pt}}$ for {\it MICROSCOPE\,} relative  to $\Delta(Q/A_r)_{\hbox{\scriptsize \,Be-Al}}$ for E\"ot-Wash. 
 \vspace{4mm}
  \label{table:grandeltaeot}}
 \begin{tabular}{c}
$ \ba{|c||c|c|c|}
\hline 
&&& \\ [-2mm]
Q/A_r&\ \  \Delta(Q/A_r)_{\hbox{\footnotesize \,Be-Ti}} \ \ &\ \ \Delta(Q/A_r)_{\hbox{\footnotesize \,Be-Al}} \ \ 
& \ -\, \dis \frac{\Delta(Q/A_r)_{\hbox{\scriptsize \,Ti-Pt}}}{\Delta(Q/A_r)_{\hbox{\scriptsize \,Be-Al}}} \ %%%& \ \ \ \ {\cal J} \ \ \ \
\\ [4mm]  \hline
&&& \\ [-3mm]
B/A_r&\ \ -\,.00242\ \  & -\,.00203  &   \ \ .39  %%%&  \  \, 5.7
\\ [.5mm]
L/A_r & \ \ -\,.01577\ \  & -\,.03797& 1.50    %%%&   22
\\ [.5mm]
\ (B-L)/{A_r} \ &  \ \ \ .01333  &\  \ \  .03593   &   1.57 %%% &   23
\\ [.5mm]
(B+L)/{A_r} &\ \  -\,.01819\ \ & \, -\,.04000      &  1.45   %%%&   21
\\ [.5mm] \ \ (3B+L)/A_r\ \  &\ \  -\,.02303 \ \ & \ -\,.04406   &  1.35   %%%&   20
\\ [.5mm] \ \ (B-2L)/A_r\ \  &\ \  \ .02910  & \ \ \ \,.07390   &    1.53  %%%&   22
\\ [1.5mm]
\hline
\ea $
\end{tabular}
\vspace{1.5mm}
\end{table}

\vspace{-1mm}

Before {\it MICROSCOPE}, the E\" ot-Wash experiment \cite{adel1,adel2} led to the most significant limits on the validity of the equivalent principle \cite{adel3,adel4},
\be
\delta_{\hbox{\footnotesize Be,Ti}}= (.3\pm 1.8)\times 10^{-13}\ \ \hbox{and} \ \ 
\delta_{\hbox{\footnotesize Be,Al}}\simeq (-\,.7\pm 1.3)\times 10^{-13}\ \  (\hbox{at } 1\sigma) \ .
\ee
With  $\,\Delta(B/A_r)_{\hbox{\scriptsize Be-Ti}} \simeq -\, .00242 $  and $\,\Delta(B/A_r)_{\hbox{\scriptsize Be-Al}} \simeq -\,.00203 $ this implies
\be
\alpha_{gB\ \hbox{\footnotesize Be,Ti}}\, = (-1.24 \pm 7.44) \times 10^{-11}\ \ \hbox{and} \ \ 
\alpha_{gB\ \hbox{\footnotesize Be,Al}}= (3.45\pm 6.40)\times 10^{-11} \ .
\ee
The two results may be combined (with weights inversely proportional to  $7.44^2$ and $6.40^2$) into
\be
\framebox [5.6cm]{\rule[-.25cm]{0cm}{.7cm} $ \dis
\alpha_{gB\ \hbox{\tiny E-\!W}}\, =\, (1.5\pm 4.9) \times 10^{-11}\,.
$}
\ee
This is to be compared with the recent {\it MICROSCOPE\,} result following from $\delta_{\hbox{\scriptsize Ti-Pt}}
=(-.1\pm1.3) \times 10^{-14}$ with $\,\Delta(B/A_r)_{\hbox{\footnotesize Ti-Pt}} \,\simeq \,.00079 $, leading to
\be
\framebox [6.2cm]{\rule[-.25cm]{0cm}{.7cm} $ \dis
\alpha_{gB\ \hbox{\tiny\,MICRO}}\, = \, (-.1\,\pm \,1.6) \times 10^{-11}\,.
$}
\ee 
The improvement brought by {\it MICROSCOPE\,} over the combined E\"ot-Wash results is by a factor $\simeq \,3$ for a coupling proportional to $B$.
The improvement from the more precise measurement of $\delta$ gets somewhat decreased due to 
the lower $\,\Delta(B/A_r)_{\hbox{\scriptsize Ti-Pt}}$, 2.5  or 3 times smaller than for the E\"ot-Wash  pairs
(cf. Tables \ref{table:grandeltamic} and \ref{table:grandeltaeot}),  and by the combination  of the 
Be-Ti and Be-Al results. These results apply for a large range $\lambda$ as compared to the radius of the Earth, the  EW experiment remaining more sensitive for $\lambda$ smaller than about a few hundred km.

\vspace{2mm}

 The improvement is larger for forces acting proportionally to $L, \ B-L, \ B+L, \ 3B+L$  or $B-2L$, as 
 the corresponding $|\Delta Q/A_r|$ are all larger for {\it MICROSCOPE\,} than for E\"ot-Wash, by a factor $\simeq 1.3$ to 1.6 for  Be-Al, and $ 2.6$ to 4.2 for 
 Be-Ti (cf. Tables  \ref{table:grandeltamic} and \ref{table:grandeltaeot}). The E\"ot-Wash constraints now come
 mainly from the Be-Al results, as understood from Table \ref{table:grandeltaeot}. To keep things simple let us illustrate this for a coupling to $B-L$, taking first into consideration the Be-Al result. The improvement factor 
 may be conservatively estimated as 
 \be
\label{il}
{\cal I}_{B-L}\,=\,\left[\,\frac{\hbox{E\"ot-Wash uncertainty on} \ \delta_{\hbox{\scriptsize Be-Al}}}{\hbox{{\it MICROSCOPE\,} uncertainty}}\,\simeq \,\frac{1.3\times 10^{-13}}{1.3\times 10^{-14}}\,\right]\,\times\,\frac{ |\Delta((B-L)/A_r)_{\hbox{\footnotesize Ti-Pt}}| \, \simeq \,.05625}{\Delta((B-L)L/A_r)_{\hbox{\footnotesize Be-Al}}\, \simeq \, .03593}\ \simeq \ 16\ .
\ee
Let us now consider the Be-Ti result. It is less constraining that the Be-Al one by a factor \linebreak $ [1.8 \times 10^{-13}/.01333] : 
[ 1.3\times 10^{-13}/.03593]\simeq 3.73$, thus weighting only for about 1/15 in the combination of the 
Be-Al and Be-Ti results.
Taking it into consideration brings the above improvement factor ${\cal J}_{B-L}$ in (\ref{il})
down from about 16 to about 15.
The situation is similar for the other couplings to $L,\,B+L,\,3B+L$ or $B-2L$.
 \ghost{
 For a coupling to $L$ the improvement factor (for 1$\sigma$ results), close to 15 for the E\"otv\"os parameters, 
is now increased by a factor $\simeq $ 1.5 from the larger  $|\Delta L/A_r|$ in {\it MICROSCOPE\,},
leading to an overall improvement factor $\,\simeq \,22$:
\be
\label{il}
{\cal I}_L\,=\,\left[\,\frac{\hbox{E\"ot-Wash limit on} \ \delta_{\hbox{\scriptsize Be-Al}}}{\hbox{MICROSCOPE limit}}\,\simeq \,\frac{2\ 10^{-13}}{1.4\ 10^{-14}}\,\right]\,\times\,\frac{\ \Delta(L/A_r)_{\hbox{\footnotesize Ti-Pt}} \ \simeq .05704}{|\Delta(L/A_r)_{\hbox{\footnotesize Be-Al}}| \simeq  .03797}\ \simeq \ 22\ .
\ee
It is almost the same for couplings to $B-L,\,B+L,\ 3B+L$ and $B-2L$, with improvement factors $\,{\cal I}\simeq $ 23, 21, 20 and 22, respectively, when comparing {\it MICROSCOPE\,} with the Be-Al E\"ot-Wash experiment, more precise than with Be-Ti in these cases (cf. Table \ref{table:grandeltaeot}).
Altogether the first {\it MICROSCOPE\,} results provide a gain in sensitivity on Equivalence Principle tests  by a factor $\simeq \,5$ for a coupling to $B$, to about 20 to 23 for the other couplings  involving $L$.
}

 \vspace{2mm}

 To express reliably absolute limits as we did in (\ref{limalphab}), however,  we need to define more precisely what we mean by ``relative intensity of the new force, with respect to gravity''.
 While this is easy for an effective coupling to $B$ for which protons and neutrons play similar roles,  it must be made more precise in the other situations.
 Let us thus discuss  the possible couplings of such a new force.

\section{Gauge symmetry, \vspace{2mm} \ grand unification
\hbox{\ \ \ \ \ \ and the couplings of a new force}}

New long-range forces may be associated with the exchanges of spin-1 or (possibly dilatonlike) spin-0 particles.
When looking for such forces we need to know, or imagine, on which quantity $Q$ the new force is supposed to act.
This may not be easy, especially for a  spin-0 induced  force \cite{berge,peccei,damour1,damour2,chamel,damour3}.  One also has to justify for having a massless or quasimassless spin-0 particle. For a spin-1 mediator, however, this can follow simply from gauge symmetry, which also
provides useful constraints on the couplings of a massless, or extremely light, spin-1 boson $U$. The expected structure of its couplings may then be discussed in connection with the other fundamental gauge interactions, weak, electromagnetic and strong, and their possible grand unification.

\vspace{2mm}

Extra-$U(1)$ gauge groups, possibly with a very light  and very weakly coupled new gauge boson $U$,  were considered very early within supersymmetric theories \cite{plb77,plb80}. This led us to discuss extensions of the standard model to
$SU(3)\times SU(2)\times U(1)_Y\times\,\hbox{extra-}U(1)$ 
or, in the case of grand unification,
$SU(5)\times U(1)$ \cite{plb86,plb89}.
After electroweak symmetry breaking leading to possible mixing effects with the photon and the $Z$,  the new neutral gauge boson $U$  generally couples to a linear combination of the baryonic, leptonic and electric charges,
$B,\ L$ and  $Q_{\rm el}$
\,\footnote{An axial part in the current may also exist when the spin-0 BEH sector includes two electroweak doublets at least, with different gauge quantum numbers \cite{plb86,plb89}. But it is experimentally very constrained, including by the possible production of a very light longitudinal $U$ boson behaving as a quasimassless axionlike pseudoscalar, even if this one can be made almost ``invisible'' using a large spin-0 singlet v.e.v.  \cite{plb80,newint}.}. Its couplings may be expressed 
in comparison with the photon coupling as 
\vspace{-2mm}
\be
\label{defeps0}
(\,\epsilon_B\,B+\epsilon_L\,L+\epsilon_{Q_{\rm el}}\,Q_{\rm el}\,)\,e\,.
\ee
The $U$ boson appears as {\em a generalized ``dark photon''} \,also coupled to a linear combination of $B$ and $L$, or $B-L$ \cite{plb86,plb89,newint,epjc}, which is of crucial importance for Equivalence Principle tests.

\vspace{2mm}

The test masses being neutral to avoid parasitic electro\-magnetic effects, any term proportional to $Q_{\rm el}$ in  (\ref{defeps0}) leads to opposite contributions from protons and electrons, of no effect here.
We can then omit the term proportional to $Q_{\rm el}$ in (\ref{defeps0}), and  get in practice an effective coupling  to
\be
\label{defeps20}
(\,\epsilon_B\,B+\epsilon_L\,L\,)\,e\ \ \ \left\{\ \ba{c} \hbox{with  a {\em fundamental origin\,} for  a spin-1 gauge boson}\ U\,,   \vspace{1mm}\\ 
 \hbox{or as an element of a {\em phenomenological parametrization\,}}
 \vspace{0mm}\\
 \hbox{for a spin-0 induced force\,.}  \ea \right.
\ee

\vspace{1mm}
\noindent
Such a coupling is also equivalent, for neutral matter with equal numbers of protons and electrons,  to a coupling to $(\,\epsilon_B\,(Z\!+\!N)+\epsilon_L\,Z\,)\,e\,$. 
A spin-1 induced force coupled as in (\ref{defeps0},\ref{defeps20}) to a conserved (or almost conserved) quantity is thus expected to have additivity properties. \,This is in contrast with a spin-0 mediated force, for which 
the possible presence of terms proportional to $Z$ and $N$ in the effective couplings should be viewed mainly as an element of {\em a phenomenological parametrization}; other contributions are generally expected, possibly involving the electrostatic and chromostatic energies \cite{damour3,berge}.  
 
\vspace{2mm}
 In the context of grand unification we have considered long ago a $SU(5)\,\times $ extra-$U(1)$ gauge group, spontaneously broken by the vacuum expectation value (v.e.v.) of the neutral component of one  (or possibly several) spin-0 BE-Higgs quintuplet(s) $\varphi$ 
 ($<\!\!\varphi^0\!\!>\ =v/\sqrt 2$ with $v\simeq 246$ GeV) into  a $SU(4)_{\rm es}\times U(1)_U$ subgroup. At the same time $SU(5)$ is spontaneously broken into $SU(3)\times SU(2)\times U(1)_Y$ in the usual way (e.g. through a spin-0  adjoint v.e.v.), leaving a $SU(3)_{\rm QCD} \times U(1)_{\rm QED}\times U(1)_U$ gauge symmetry \cite{plb89}.
After mixing effects between neutral gauge bosons, the coupling of the resulting massless (or extremely light)
gauge boson $U$ gets expressed at  the grand unification scale as 

\vspace{-5mm}
\be
\label{epsgut}
\epsilon_{B-L}\,(B-L-\,\hbox{\small$\dis \frac{1}{2}$}\ Q_{\rm el})\ e\ .
\ee 
It preserves an $\,SU(4)_{\rm es}$ {\it electrostrong symmetry\,}  including $SU(3)_{\rm QCD}\times U(1)_{\rm QED}$, unifying directly electromagnetic with strong interactions at very high energies, and the photon with the eight gluons
\cite{epjc,fayetmicro}.  

The surviving $U(1)_U$ generator $Q$, commuting with this electrostrong $SU(4)_{\rm es}$, is such that
%%\vspace{-1mm}
\be
Q=B-L\,-\hbox{\small$\dis \frac{1}{2}$}\,Q_{\rm el} =  -\,\hbox{\small$\dis \frac{1}{2}$}\  \ \,\hbox{for the vectorial antiquartet}\
\left(\,\ba{c}
\bar d\vspace{.5mm}\\ \bar d\vspace{.5mm}\\ \bar d\vspace{.3mm}\\ e \ea\,\right)_{\!\!L+R}\!\!\!\!,
  \ \ \ 0 \ \ \hbox{for the sextet}\ 
 \left(\ba{rrrc} 0 & \bar u & \!\!-\bar u & -u\vspace{.5mm}\\
\!\!-\bar u & 0 &\bar u & -u \vspace{.5mm}\\
\bar u &\! \!-\bar u & 0 & -u \vspace{.5mm}\\
u &  u & u & \ \ 0
\ea \right)_{\!\!L} \!\!.\ 
\ee
Expression (\ref{epsgut}) of the $U$ coupling, valid at the grand-unification scale, 
gets modified at lower energies into 
$\,\epsilon_{B-L}$ $(B-L-\,\frac{4}{5}\cos^2\theta \,Q_{\rm el})\,e$ \cite{plb89}, with $\cos^2\theta$ increasing from 5/8 at the grand-unification scale up to about .762 at low energies. It is then equal to
\be
\label{defeps02gut}
\framebox [5cm]{\rule[-.25cm]{0cm}{.7cm} $ \dis
\epsilon_{B-L}\,(B-L\,-.61\ Q_{\rm el})\ e\ .
$}
\ee
With $B$ and $L$ occurring through $\,B\!-\!L$ (which tends to remain conserved or approximately conserved  in this context, allowing  e.g. for the decay $p\to \pi^0 e^+$), the effective coupling (\ref{defeps20}) reduces to
\be
\label{defepsb-l}
\epsilon_{B-L}\,(B-L)\,e\,.
\ee

\vspace{1mm}
In another approach one can consider a left-right symmetric theory with a Pati-Salam \cite{ps} gauge group, extended to $[\,SU(4)_{L+R}
\times SU(2)_L\times SU(2)_R\,]\,  \times $ extra-$U(1)$ \cite{plb89}, acting on the quark and lepton quartets
%%\vspace{-3mm}
\be
\left(\,\ba{cccc}  \vspace{-5mm} \\ u&u&u&\,\nu \\ d & d & d & \,e \ea\right)_{\!L+R} 
\ee
with $\,3B+L=1$. This leads to an  extended electroweak gauge group 
\be
SU(2)_L\times SU(2)_R \times  U(1)_{B-L}\times \hbox{extra-}  U(1)_{\,3B+L}\,.
\ee
The $U$ boson is then coupled to 
\vspace{-2mm}
\be
\epsilon_{\,3B+L}\,(3B+L)\,e\ ,
\ee
as long as it does not mix with the other neutral gauge bosons.
Otherwise the $3B+L$ current combines with the other neutral currents. With the vector part in $J^\mu_Z$ a linear combination of $B\!-\!L$ and electromagnetic currents, we return  as usual, for the vector part in the coupling, to a linear combination of $B$ and $L$ with  the electric charge  as in (\ref{defeps0}).

\section{\boldmath Limits on the strength \vspace{2mm} of a new force \hbox{compared to gravity}}
 \vspace{-4mm}
 \bc
{\bf   (\ {\boldmath $|\bar\alpha_{g}|< $ 
{\normalsize $2.2 \times 10^{-13}\,$}}\ {\it\em (at 2$\sigma$)}\ 
\,for a coupling to \boldmath  $\,B\!-\!L$\,)}
\ec

\vspace{-6mm}

\subsection{E\"otv\"os parameter}

We shall thus be concerned with a force acting effectively on a charge $Q$ linear combination of $B$ and $L$ as in (\ref{defeps0}), and more specifically  $B\!-\!L\,$ in the case of grand unification as in (\ref{defeps02gut},\ref{defepsb-l}), even if this  is well motivated only for a  spin-1 induced force.
For spin 0 this may be viewed, at best, as a phenomenological description, next to other possible contributions to the couplings \cite{peccei,damour1,damour2,damour3,chamel,berge}.
With
\be
\label{vbl}
V_{BL}(r)\,=\,\pm \ \frac{e^2}{4\pi\epsilon_\circ}\,(\epsilon_BB\!+\!\epsilon_LL)_a\ (\epsilon_BB\!+\!\epsilon_LL)_i 
\ \,\frac{e^{-r/\lambda}}{r}
\,=\,\pm \ \epsilon_Q^2\  \alpha\ Q_aQ_i 
\ \frac{e^{-r/\lambda}}{r}\,,
\ee
expression (\ref{defalphab}) of the potential gets modified into
\be
\label{defalpha20}
V(r) \,=\, -\,\frac{G_N\,m_a m_i}{r}\ \left[\ 1+\alpha_{gQ} \,\left(\frac{Q}{A_r}\right)_a\,\left(\frac{Q}{A_r}\right)_i \ e^{-r/\lambda}
\ \right]\ ,
\ee
with
\be
\label{aleps2}
\framebox [7.8cm]{\rule[-.35cm]{0cm}{.9cm} $ \dis
\alpha_{gQ} \,=\, \mp\ \frac{\alpha}{G_N\,{\rm u}^2}\  \,\epsilon_Q^2\  \simeq\ \mp \ 1.2536\times 10^{36}\ \epsilon_Q^2\,,
$}
\ee
as expressed  in (\ref{aleps}) with $Q=B$.

\vspace{2mm}
The E\" otv\" os parameter 
$
\delta_{1 2}=2\,(a_1-a_2)/(a_1+a_2)$ $\simeq (a_1-a_2)/g\,\simeq \, \delta_1-\delta_2
$
measures the relative difference in the observed accelerations $a_i$ of two test masses ``freely-falling'' toward the Earth.
For a force of range $\lambda$ sufficiently large compared to the Earth radius this leads to the estimate
\be
\delta_{12}\ \simeq\ \alpha_{g Q}\  \left(\frac{Q}{A_r}\right)_{\oplus}\Delta\! \left(\frac{Q}{A_r}\right)_{\!12}
 \ \simeq\ \mp \ 1.2536\times 10^{36}\, \left(\frac{Q}{A_r}\right)_{\oplus}\Delta\! \left(\frac{Q}{A_r}\right)_{ \!12} \, \epsilon_Q^2\ ,
\ee
reducing to (\ref{aleps},\ref{deltazero}) for $Q\!=\!B$.
\,Such an $\alpha_g$ is often considered as representative of the intensity of the new force, if long ranged, as compared to gravity. This is, however, misleading. It may already be understood as  $\alpha_{gQ}$, as defined in (\ref{defalpha20},\ref{aleps2}), depends on the normalization chosen for the charge $Q$ while the relative intensity of the new force, compared to gravity, should 
be independent of it. This leads us to define an absolute normalization for $Q$, redefined into $\bar Q$ normalized to 1 for an ``average nucleon", i.e. so that $p+e+n$ has $\,\bar Q=B=2$.

\subsection{\boldmath Defining {\normalsize $\,\bar\alpha_g$} \,for an average nucleon}

To illustrate this we can ask whether $\alpha_{gL}$, as defined from (\ref{defalpha20}), represents, at least approximately, the intensity of a new force effectively coupled to $L$, compared to gravity.  This is true between two protons with their accompanying electrons, 
but such a force does not act on neutrons. For similar numbers of protons and neutrons  the force gets reduced by an extra factor $\simeq 4$ as compared to gravity, and is better represented by  $\bar \alpha_{gL}= \alpha_{gL}/4\,$ than by the original $\alpha_{gL}$.
This is the same for a force coupled to $B\!-\!L=N$ acting effectively only on neutrons \cite{plb86,plb89},
 better represented by  $\bar \alpha_{g\,B-L}= \alpha_{g \,B-L}/4\,$ than by the original $\alpha_{g\,B-L}\,$.

 \vspace{2mm}

We shall thus define  $\bar\alpha_g$ by referring to ideal isoscalar bodies, with equal numbers of protons and neutrons. And renormalize any effective charge $Q=xB+yL$ into
$\bar Q$, equal to 1 for an average nucleon, through the redefinition 
\vspace{-1mm}
\be
Q=xB+yL\ \ \to \ \ \bar Q \,=\,\frac{2}{2x+y}\ Q\, = \ \ \hbox{e.g.}\ \left\{\ \ba{ccc}
B\,,
\vspace{.5mm}\\
2L\,,
\vspace{.5mm}\\
2\,(B-L)\ ,
\vspace{.5mm}\\
2\,(B+L)/{3}\ ,
\vspace{.5mm}\\
{2\,(3B+L)}/{7}\ .
\ea
\right.
\ee

\vspace{-2mm}

\noindent
$\bar Q $ is normalized to \,
\be
\bar Q\,(\,p+e+n\,)\,=\,B\,=\,2\,.
\ee
Expressing  $\,\bar\alpha_g\,\bar Q_a\bar Q_i \equiv\alpha_g\,Q_aQ_i\,$ for the new contribution  (\ref{vbl}) to the potential (\ref{defalpha20})
we get, for couplings to $B,\,L,\,B-L,\,B+L$ or $3B+L$, 
\be
\label{alphabar}
\bar \alpha_{gQ}\,=\,\left(x+\frac{y}{2}\right)^2 \alpha_{gQ}\,= \left\{ \ba{cccc}
&\alpha_{gB} & \hbox{(same action on $p$ and $n$)}
\vspace{.5mm}\\
(1/4)&\alpha_{gL} &  \hbox{(acts only on $p$)}
\vspace{.5mm}\\
(1/4) &\alpha_{g\,B-L}&   \hbox{(acts only on $n$)}
\vspace{.5mm}\\
(9/4)&\alpha_{g\,B+L}&\   \hbox{(acts on average nucleon 3/2 as much as on $n$)}\!\!\!\!\!\!\!
\vspace{.5mm}\\
(49/4) &\alpha_{g\,3B+L}\ .\!\!\!&&\ \ \ \ \ 
\ea
\right.
\ee

 \begin{table}
 \caption{\ Limits on the relative strength $\bar \alpha_g$ of a long-range force coupled to $Q$. We use 
$|\delta| < 2.5 \times 10^{-14}$ %% for $\delta<0$, and  $< 2\times 10^{-14}$  for $\delta>0$, 
at $2\sigma$ \cite{micro},
with   $({Q}/{A_r})_{\oplus}$ and $\Delta ({Q}/{A_r})_{\hbox{\scriptsize \,Ti-Pt}}$ from \cite{fayetmicro}. 
For a coupling to $N\!-\!Z$\,,\,  
 $|\alpha_{g\,B-2L}|< 8.4\times 10^{-12}$  represents the relative strength of the new force between two nucleons.
 }
 \label{table:lim}
 \vspace{3.5mm}
$
\ba{|c||r|r|lcc||c|c|}
\hline 
&&&&&& \\ [-3mm]
Q &\ \dis  \left({Q}/{A_r}\right)_{\oplus} \,&\dis \  \Delta\! \left({Q}/{A_r}\right)_{\hbox{\scriptsize \,Ti-Pt}} \ && \!\!\!\!\!\!\!\!\!\delta_{\hbox{\footnotesize Ti-Pt}}\!\!\!\!\!\!\!\!\!&  &
 \ \ \ \lim  |\bar\alpha_g | \ \, (2\sigma)\ 
\\ [2mm]
\hline
&&&&&&\\ [-2.5mm]
B & 1.0008  \ \ & .00079 \ \ \ \ \  &\ \ \ \ .00079\ \ \,\alpha_{gB} &\!=\!&\! .00079\  \ \bar\alpha_{gB}\ \,
& \ \ 3.2\times 10^{-11}\ \ 
\\ [1mm]
L& .4870 \ \ &  .05704  \ \ \ \ \ & \ \ \ \ .02778 \ \ \,\alpha_{gL}  &\!\simeq\!&  \!  .11111\ \ \bar  \alpha_{gL}\ \,
& 2.3\times 10^{-13}
\\ [1mm]
B\!-\!L &.5138 \  \ &   -\,.05625 \ \ \ \ \ &\  -\,.02890 \ \ \alpha_{g\,B-L}\!\! \!\!&\!\simeq\!&\!\!-\,.11560\ \bar \alpha_{g\,B-L}\ \,
& 2.2\times 10^{-13} 
\\ [1mm]
\ B\!+\!L \ &1.4878\ \ & .05783  \ \ \ \ \ & \ \ \ \ .08604\  \ \alpha_{g\,B+L} \!\!&\!\simeq\!& \ \ .03824\ \bar \alpha_{g\,B+L} \ \,
& 6.7\times 10^{-13}
\\ [1mm]
\ 3B\!+\!L \ & 3.4894\ \ & .05941  \ \ \ \ \ &\ \ \ \ .20731\  \ \alpha_{g\,3B+L}  &\!\simeq\!& \ \ \,.01692\ \bar \alpha_{g\,3B+L} \  
& 1.5\times 10^{-12}
\\ [1mm]
\ B\!-\!2L \ & 0.0268\ \ & -\,.11329  \ \ \ \ \ &\  -\, .00304\  \ \alpha_{g\,B-2L} \!\!&&  /
& / 
\\ [2.5mm]
\hline
\ea
\vspace{1mm}
$
\end{table}

\vspace{-2mm}

\subsection{\boldmath Limits on  \normalsize $\,|\bar\alpha_g|$}

\vspace{-2mm}

For a long-range force one has
\be
\bar\alpha_g\,=\,\frac{\hbox{effective new force between average nucleons}}{\hbox{gravitational force between average nucleons}}
\ee
(treating apart a coupling to isospin $I_3=(Z-N)/2=L-B/2$, for which $\bar\alpha_g$ would vanish).
The resulting limits on $\alpha_g$ and $\bar\alpha_g$ are obtained as
\be
\label{limalphag}
\lim |\alpha_g|\,=\,\frac{|\lim \delta|}{({Q}/{A_r})_{\oplus}  \  |\Delta ({Q}/{A_r})|}\ ,
\ee 
and
\vspace{-2mm}
\be
\label{limbar}
\lim |\bar\alpha_g|\ =\ \frac{|\lim \delta|}{ \underbrace{({\bar Q}/{A_r})_{\oplus} }_{\hbox{\small $\simeq \,1$}}\  |\Delta ({\bar Q}/{A_r})|}
\ \simeq\ \left\{
\ba{ccc}
 \dis\frac{|\lim \delta |}{|\Delta ({B}/{A_r})|}\ \ \ \ \ \ \ &\simeq &\ \  3.2\times 10^{-11}\ ,
 \vspace{2mm}\\
  \dis\frac{|\lim \delta |}{|\Delta ({2L}/{A_r})|}\,\times\,
  \left\{\ba{c}
  1\vspace{0mm}\\
    1 \vspace{0mm}\\
      3 \vspace{0mm}\\
        7 \vspace{0mm}\\
  \ea\right\} &\simeq &
  \left\{ \ba{c}
  2.3\times 10^{-13}\ ,  \\
    2.2\times 10^{-13}\ ,  \\
      6.7\times 10^{-13} \ , \\
        1.5\times 10^{-12} \ , \\
\ea\right.
\ea\right.
\ee 

\vspace{2mm}

\noindent
as given in Tables \ref{table:lim} and \ref{table:lim2} (with $(\bar Q/A_r)_\oplus$ close to 1). This also shows the interest in most cases  of trying to maximize  $| \Delta (L/A_r)|$
or $| \Delta (B\!-\!L)/A_r)|$, preferentially to $ |\Delta (B/A_r)|$. In particular, we have
\be
\framebox [6.2cm]{\rule[-.25cm]{0cm}{.7cm} $ \dis
|\bar\alpha_{g\,B-L}|\ <   \ 2.2 \times 10^{-13}\ \ (\hbox{at} \ 2\sigma)
$}
\ee
for a coupling to $\,B\!-\!L\,$.

 \begin{table}
 \caption{\ 
Limits on $\bar \alpha_g$ as in Table \ref{table:lim}, obtained directly as in (\ref{limbar}). 
They are close to $\lim \delta\,/ \Delta (B/A_r)$ and  $[\,\lim \delta\,/ \Delta (2L/A_r)\,]\times  (1,\,-1,\,3\, $ or 7). 
We use 
$|\delta| < 2.5 \times 10^{-14}$ at $2\sigma$  %%%(or $\,< 1.3 \times 10^{-14}$   at $1\sigma$)
 \cite{micro}. \,The $1\sigma $ limits on $|\bar \alpha_g|$ are \hbox{2 times smaller.}}
 \label{table:lim2}
 \vspace{4mm}
$\ba{|c|c||r|c|c||c|}
\hline 
&&&&& \\ [-2.5mm]
Q &\,\bar Q& \ \ (\bar Q/A_r)_{\oplus} \ &\dis \  \,\Delta \,(\bar Q/A_r)_{\hbox{\scriptsize Ti-Pt}} \ & \!\!\!\!\!\!\!\!\!\delta_{\hbox{\scriptsize Ti-Pt}}\!\!\!\!\!\!\!\!\!&
\ \ \ \lim  |\bar\alpha_g | \ \, (2\sigma)\ 
\\ [2mm]
\hline
&&&&&\\ [-2mm]
B &B& 1.0008  \ \ \ &\ \   .00079 \ \    & \ \  .00079\  \ \bar\alpha_{gB}\ \ \ \
&   \ \ 3.2\times 10^{-11}\ \ 
\\ [1mm]
L&2L& .9740 \ \ \ & \ \  .11408  \ \ &  \ \  .11111\ \ \bar  \alpha_{gL}\ \ \ \
&  2.3\times 10^{-13}
\\ [1mm]
B\!-\!L &2\,(B-L)& 1.0276 \  \ \ &    -\,.11250 \ \ \ & -\,.11560\ \bar \alpha_{g\,B-L}\ \
& 2.2\times 10^{-13} 
\\ [1mm]
\ B\!+\!L \ & 2\, (B+L)/3&.9919\ \ \ &\ \  .03855  \ \ & \   \,.03824\ \bar \alpha_{g\,B+L}\
&   6.7\times 10^{-13}
\\ [1mm]
\ 3B\!+\!L \ & \ 2\,(3B+L)/7\ &.9970\ \ \ & \ \ .01697  \ \ & \   \  \, .01692\ \bar \alpha_{g\,3B+L}\ 
&   1.5\times 10^{-12}
\\ [3mm]
\hline
\ea
\vspace{4mm}
 $
\end{table}

\vspace{2mm}

In doing so we left aside the case of  a new force coupled to $B-2L=N-Z=\,-\,2 \,I_3$, acting effectively oppositely on protons and neutrons, with no action on an ``average nucleon''.
 $\bar \alpha_g$, vanishing, is no longer relevant.
We can then simply use $\alpha_{g\,B-2L}$ as a measure of the relative intensity of the new force, if long ranged, between two protons (with their electrons), or two neutrons, or a proton and a neutron, relative to gravity.
From  eq.\,(\ref{limalphag}) and Table \ref{table:lim} we obtain
\be
\label{limalphag2}
|\alpha_{g\,2B-L}| \,<\,
\frac{2.5\times 10^{-14}}{.0268\times .11329}\,\simeq\,8.4\times 10^{-12}\ .
\ee 
This limit is less restrictive than the ones on $|\bar\alpha_{gQ}|$ for effective charges involving $L$, as seen in Tables \ref{table:lim2} and \ref{table:lim3}, owing to  the small value of $\,((B-2L)/A_r)_\oplus\simeq .0268$\,.

\section{Comparison with electromagnetism}

\subsection{\boldmath Limits on \normalsize $\,\epsilon_Q$}

\vspace{-1mm}

\ghost{%%%%%%%%%%%%%%%%%%%%%%%%%%
 \begin{table}[t]
 \caption{\  2$\sigma$ limits on $\bar \alpha_g, \,\alpha_g$ and $\epsilon$, with
$ \bar\alpha_g=\alpha_g \times (1,\,1/4,\,1/4,\,9/4,\,49/4)$,
and $\alpha_g\simeq \mp \ 1.2536\times 10^{36}
\,\epsilon^2$. The limits on $|\bar \alpha_{gQ}|$ are close to $2.5\times 10^{-14}/\, |\Delta(\bar Q/A_r)|$,
\,with $|\epsilon_Q| < 1.43\times 10^{-25}/ 
\sqrt{(Q/A_r)_\oplus\! \times |\Delta(Q/A_r)|}$ \,(cf. Table \ref{table:lim} and eqs.\,(\ref{deltaeps0},\ref{deltaeps})).
}
 \label{table:lim3}
 \vspace{2mm}
$\ba{|c||c|c|r|}
\hline 
&&& \\ [-3mm]
\ \ \ \  \ \ Q \ \ \ \ & \ \ \ \  |\,\bar\alpha_{g}|  < \ \ \  \ & \ \ \ \
 |\,\alpha_{g}\, | < \ \ \ \ & \ \ \ \ \ \  |\,\epsilon\,| < \ \ \ \  \ \ 
\\ [1.5mm]
\hline
&&&\\ [-2.5mm]
B & 3.2\times 10^{-11}  &  3.2\times 10^{-11}  &  \ 5\   \times 10^{-24} \ \ 
\\ [1mm]
L& 2.3\times 10^{-13} &  9.2\times 10^{-13} & .86\times 10^{-24}\ \ 
\\ [1mm]
B\!-\!L & 2.2\times 10^{-13} & 8.8\times 10^{-13}  & .84\times 10^{-24}\ \ 
\\ [1mm]
\ B\!+\!L \ &  6.7\times 10^{-13} &  3\ \ \times 10^{-13} & \ .49\times 10^{-24}\ \ 
\\ [1mm]
\ 3B\!+\!L \ &  1.5\times 10^{-12} & 1.2 \times 10^{-13} & \ .32\times 10^{-24}\ \ 
\\ [1mm]
\ B\!-\!2L \ &  /  &  8.4  \times 10^{-12} & 2.6 \, \times 10^{-24}\ \ 
\\ [2.5mm]
\hline
\ea
\vspace{3mm}
 $
\end{table}
}%%%%%%%%%%%%%%%%%%%%%%%%%

For a new force acting effectively proportionally to a charge $Q$ linear combination of $B$ and $L$, with an effective coupling 
$\,\epsilon_Q Q\,e=(\epsilon_B B+\epsilon_L L)\,e\,$ with  $\,e=\sqrt {4\pi\alpha}\simeq \,.3028$, the potential
and modified Newton potential are given by eqs.\,(\ref{vbl},\ref{defalpha20}), with
$ \alpha_{gQ}\,\simeq\,\mp\ 1.2536\times 10^{36}\ \epsilon_Q^2\,.$
This leads for a long-range force (as compared to the Earth radius) to the E\" otv\"os parameter \cite{fayetmicro}
\be
\label{deltaeps0}
\ba{ccl}
\delta_{12}\,=\,\alpha_{gQ}\ (Q/A_r)_\oplus \ \Delta(Q/A_r)_{12}&=& \mp\ 1.2536\times 10^{36}\ 
\left(\epsilon_B \hbox{\small$\dis\frac{B}{A_r}$}+ \epsilon_L  \hbox{\small$\dis\frac{L}{A_r}$}\right)_\oplus \  
\left(\epsilon_B \,\Delta\hbox{\small$\dis\frac{B}{A_r}$}+\epsilon_L \, \Delta \hbox{\small$\dis\frac{L}{A_r}$}\right)_{12} 
\vspace{2mm}\\
&\simeq & \underbrace{\mp\ 1.2536\times 10^{36}\ \epsilon_Q^2}_{\hbox{$\alpha_{gQ}$}}\ \
(Q/A_r)_\oplus \ \Delta(Q/A_r)_{12}\,,
\ea
\ee
with $(B/A_r)_\oplus \simeq 1.0008$ and $(L/A_r)_\oplus \simeq .4870$.
The upper and lower signs are  for spin-1 and spin-0 mediators, respectively.
With  $(Q/A_r)_\oplus$  and $\Delta(Q/A_r)_{\hbox{\scriptsize Ti-Pt}}$ in Table \ref{table:lim} this reads
\,\footnote{The 1.00 (rather than .99) in the expression of $\delta_B$ originates from a more precise estimate of $\Delta(B/A_r)_{\hbox{\scriptsize Ti-Pt}}$, above .00079.}
\be
\label{deltaeps}
\delta_{\hbox{\scriptsize Ti-Pt}}\,\simeq\,
\left\{\ \ba{lcc}
\mp\   1.00       & \!  \times\  10^{33}  \!  & \!  \epsilon_B^2\,,
\vspace{1mm}\\
\mp\       3.482  & \!   \times \  10^{34} \!   & \!  \epsilon_L^2\,,
\vspace{1mm}\\
\pm\      3.623    &  \!  \times \ 10^{34}  \!  & \!   \epsilon_{B-L}^2\,,
\vspace{1mm}\\
\mp\   1.079   &  \!   \times  \  10^{35}   \!   & \!  \epsilon_{B+L}^2\,,
\vspace{1mm}\\
\mp\    2.599   &  \!  \times  \  10^{35}   \!  & \!    \epsilon_{3B+L}^2\,,
\vspace{1mm}\\
\mp\   3.811    &   \!   \times \ 10^{33} \!    & \!   \epsilon_{B-2L}^2\,.
\ea\right.
\ee
The resulting limits on $|\epsilon_Q|$ are given in Table \ref{table:lim3}, including
\be
\framebox [9.2cm]{\rule[-.25cm]{0cm}{.7cm} $ \dis
|\,\epsilon_B| < \ 5\times 10^{-24}\,,\ \  |\,\epsilon_L|\ \ \hbox{or}\ \ |\,\epsilon_{B-L}|\ < \ .86\times 10^{-24}\ ,
$}
\ee
and 

\vspace{-5mm}
\be
|\,\epsilon_{B-2L}|\ < \ 2.6\times 10^{-24}
\ee
for an effective coupling to $B\!-\!2L=N\!-\!Z=-\,2\,I_3\,$, \,involving isospin. This limit is weaker due the partial cancellation effect between the effective contributions of protons and neutrons in the Earth.

 \begin{table}[t]
 \caption{\  2$\sigma$ limits on $\bar \alpha_g, \,\alpha_g$ and $\epsilon$, with
$ \bar\alpha_g=\alpha_g \times (1,\,1/4,\,1/4,\,9/4,\,49/4)$,
and $\alpha_g\simeq \mp \ 1.2536\times 10^{36}
\,\epsilon^2$. The limits on $|\bar \alpha_{gQ}|$ are close to $2.5\times 10^{-14}/\, |\Delta(\bar Q/A_r)|$,
\,with $|\epsilon_Q| < 1.43\times 10^{-25}/ 
\sqrt{(Q/A_r)_\oplus\! \times |\Delta(Q/A_r)|}$ \,(cf. Table \ref{table:lim} and eqs.\,(\ref{deltaeps0},\ref{deltaeps})).
}
 \label{table:lim3}
 \vspace{2mm}
$\ba{|c||c|c|r|}
\hline 
&&& \\ [-3mm]
\ \ \ \  \ \ Q \ \ \ \ & \ \ \ \  |\,\bar\alpha_{g}|  < \ \ \  \ & \ \ \ \
 |\,\alpha_{g}\, | < \ \ \ \ & \ \ \ \ \ \  |\,\epsilon\,| < \ \ \ \  \ \ 
\\ [1.5mm]
\hline
&&&\\ [-2.5mm]
B & 3.2\times 10^{-11}  &  3.2\times 10^{-11}  &  \ 5\   \times 10^{-24} \ \ 
\\ [1mm]
L& 2.3\times 10^{-13} &  9.2\times 10^{-13} & .86\times 10^{-24}\ \ 
\\ [1mm]
B\!-\!L & 2.2\times 10^{-13} & 8.8\times 10^{-13}  & .84\times 10^{-24}\ \ 
\\ [1mm]
\ B\!+\!L \ &  6.7\times 10^{-13} &  3\ \ \times 10^{-13} & \ .49\times 10^{-24}\ \ 
\\ [1mm]
\ 3B\!+\!L \ &  1.5\times 10^{-12} & 1.2 \times 10^{-13} & \ .32\times 10^{-24}\ \ 
\\ [1mm]
\ B\!-\!2L \ &  /  &  8.4  \times 10^{-12} & 2.6 \, \times 10^{-24}\ \ 
\\ [2.5mm]
\hline
\ea
\vspace{1mm}
 $
\end{table}

\subsection{Relations between limits}

For a better understanding let us have a look at the relations between the upper limits on $|\bar \alpha_{gQ}|$ or $|\alpha_{g Q}|$,  and $|\epsilon_Q|$.
\,The $2\sigma$ upper limit  $|\epsilon_B| < 5\times 10^{-24}$ \cite{fayetmicro} corresponds to $\,|\bar \alpha_{gB}|\equiv |\alpha_{gB}|<
3.2\times 10^{-11}$  in (\ref{limalphab})  and Table \ref{table:lim}:
\be
\label{nfgr}
\left|\,\frac{\hbox{new force}}{\hbox{gravity}}\,\right|\,\simeq\,|\bar\alpha_{gB}| = | \alpha_{gB}|\, \simeq \, 1.2536\times 10^{36} \ \epsilon_B^2\, <
\ 3.2\times 10^{-11}\  \ \Longleftrightarrow\ \   |\epsilon_B| <\ 5\times 10^{-24}\  .
\ee
The new force between two protons with their accompanying electrons,
compared to the electromagnetic force between these protons, satisfies
\be
\label{nfem}
\dis
\left|\,\frac{\hbox{new force}}{\hbox{electromagnetic force}}\,\right|\,= \ \epsilon_B^2 \ < \ 2.5\times 10^{-47}\,.
\ee
These ratios (\ref{nfgr},\ref{nfem})
involve as in (\ref{aleps})
\be
\label{emgr}
\dis
\left|\,\frac{\hbox{electromagnetic force}}{\hbox{gravity}}\,\right|\,\simeq\,\ \frac{\alpha}{G_N\,\hbox{u}^2}\,=\,\frac{|\alpha_{gB}|}{\epsilon_B^2}
\simeq \,1.2536\times 10^{36}\ .
\ee

\vspace{2mm}

More generally for couplings proportional to 
$B,\ L,\,B-L,\,B+L$ or $3B+L$,
one has \,\footnote{We may also define $ \epsilon_{\bar Q}= \epsilon_Q\times  (1,\,1/2,\,1/2,\,3/2\ \hbox{or}\  7/2)$ so that $\bar\epsilon\,\bar Q\equiv\epsilon Q$ and $\,\bar\alpha_{g Q}  \simeq \,\mp\ 1.2536\times 10^{36} \ \epsilon_{\bar Q}^2$\,, although this seems of reduced practical interest.}
\be
\bar\alpha_{g Q} \, \simeq \, \mp\ 1.2536\times 10^{36} \ \epsilon_Q^2\, \times  (1,\,1/4,\,1/4,\,9/4\ \hbox{or}\  49/4)\,
\ee
(and for a coupling to  $B-2L$, 
$
\alpha_{g \,B-2L} \, \simeq \, \mp \ 1.2536\times 10^{36} \ \epsilon_{B-2L}^2
$).
The corresponding limits, given in Table \ref{table:lim3}, are related 
\vspace*{-2mm} by 
\be
\lim |\epsilon_Q|\,\simeq \,.893\times 10^{-18}\ \lim \sqrt  {|\bar\alpha_{gQ}|}\ \times 
 \left\{ \ba{c}
  1\\
    2 \\
     2 \\
    2/3  \\
    2/7
\ea\right.
\simeq \ 
\left\{\ \ba{c}
  5\ \times 10^{-24}\,,\\
     .86\times 10^{-24}\,, \\
        .84\times10^{-24}\,, \\
       .49\times 10^{-24}\,, \\
          .32\times 10^{-24}\,.
\ea\right.
\ee

\vspace{1mm}

\section{\boldmath The \,very high energy \ $\leftrightarrow$ \ very small coupling \,connection}

\vspace{-1mm}
\bc
{\bf\boldmath A huge energy density \vspace{1.5mm}
possibly at the origin  of inflation,
\hbox{ from an extremely weak  \hbox{new interaction}}}
\ec
 \vspace{1mm}

The corresponding hierarchy between gauge couplings, by a factor $\simge 10^{24}$, may be associated within supersymmetry with a large hierarchy in energy scales by a factor $\simge 10^{12}$. Indeed in a supersymmetric extension of the standard model  with an extra-$U(1)$ gauge group, a new $U(1)$ gauge coupling $g"$, and the corresponding  $\,\xi" D"$ term \cite{fi} in the Lagrangian density, we can consider the limit 
\cite{ssm}
\be
\xi" \to \infty\ \ \hbox{i.e. \em  extremely large}, \ \ g"\to 0\ \ \hbox{i.e. \em extremely small},\ \ \hbox{with \  $\xi"g"/2= \mu_\circ^2$ \  fixed}\,.
\ee
It generates in the Lagrangian density, from the expansion of the $D"^2/2$ contribution to the potential \cite{fayetmicro} 
\vspace*{-3mm}
as
\be
\label{pot}
V"=\frac{D"^2}{2}\, = \,\frac{\xi"^2}{2}\,+\,\sum_i\ \frac{\xi"g"}{2}\,F_i\ \varphi_i^\dagger\varphi_i\,+\,...\ ,
\ee
the soft supersymmetry-breaking spin-0 mass$^2$  coefficients
\be
\label{mui}
\mu_i^2\,=\,\frac{\xi"g"}{2}\,F_i\, = \,  \mu_\circ^2 \,F_i\,.
\ee
$F_i$ denotes the extra-$U(1)$ gauge quantum numbers for left-handed chiral superfields (with
$\xi"$ possibly replaced in this expression by an effective $\xi"_{\rm eff}$).
At the same time the  $\,\xi" D"$ term in $\cal L$ induces in the potential of scalar fields (\ref{pot}) a field-independent contribution
\be
V"_{\!\!\circ}\,=\,\frac{1}{2}\ \xi"^2\,\propto\,\frac{1}{g"^2}\,,\ \ \hbox{or}\ \ \frac{1}{\epsilon^2}\ ,\ \ \hbox{huge}\,.
\ee
This huge energy density $V"_{\!\!\circ}$ originating from the $\xi"D"$ term \cite{fi} 
may be at the origin of the very rapid inflation of the early Universe, in connection with a new long-range force \cite{fayetmicro,plb84}. 
This leads in the simplest case to a rough evaluation of the very large inflation scale by $V"_{\!\!\circ} \approx \Lambda_{\rm inflation}^4$\,, corresponding to an extremely small  gauge coupling $\,g"$.

\vspace{2mm}
Beyond that, depending on the specific situation considered and on how spontaneous supersymmetry breaking is affected by inflation,
 $\,g"$  may tentatively be evaluated, assuming for simplicity $\xi"_{\rm eff}$ and $\xi"$ of similar orders of magnitude,
as  \cite{fayetmicro}
\be
\label{inf}
|\,g"|\,\approx \,|\,\epsilon\,| \,e\,\approx \,\frac{m_{\hbox{\footnotesize \,sparticle}}^2}{|\,\xi"|}\,\approx \,\left(\frac{1-10\ \hbox{TeV}}{\Lambda_{\rm inflation }}\right)^2\,.
\ee
For an effective coupling to $B-L$, as suggested by grand unification, one has
\be
\epsilon_{\hbox{\tiny $B\!\!-\!\!L$}}e\ \simeq \ -\,\frac{5}{4}\ g"\,,
\ee
with $|\epsilon_{\hbox{\tiny $B\!\!-\!\!L$}}|< .84\times 10^{-24}$ requiring an extra-$U(1)$ gauge coupling  
\be
|\,g"|\ <\ 2\times 10^{-25}\,.
\ee
As supersymmetric particles have not been found at LHC, this corresponds to an expected inflation scale larger than $\approx$ 1 to 10 TeV by about 12 orders of magnitude at least, thus $\,\simge \,10^{15}$\,--\,$10^{16}$ GeV. It may also be associated with a large gravitino mass (typically $\approx \,10^{11}$ to $10^{14}$ GeV), depending however on the details of the supersymmetry-breaking mechanism after the end of inflation.

\vspace{2mm}
Conversely an inflation scale of the order of $10^{16}$ GeV, as commonly assumed, would correspond along these lines to a very small $|\epsilon| \simle 10^{-24}$, depending on the assumptions for the symmetry-breaking mechanism, and resulting mass parameters and sparticle masses.
Such a new interaction, although extremely weak,  could still be accessible to Equivalence Principle tests.
Indeed  for the {\it MICROSCOPE\,} experiment with a spin-1 $U$ boson coupling to $B\!-\!L$,  the E\"otv\"os parameter may be
\vspace{1mm}
\be
\ba{ccccccl}
\delta_{B-L\,}(\hbox{Ti-Pt})\,&\simeq &- \,.0289\ \alpha_{g\,B-L}&\simeq &  3.623 \times 10^{34}\ \epsilon_{B-L}^2 &\simeq &
\ \ \ \ .617\times 10^{36}\ g"^2 
\vspace{3mm}\\
&\simeq &\dis 2.5\times 10^{36}\ \frac{\mu_\circ^4}{\xi"^2}&\simeq &
\dis 1.23\times  10^{36}\  \,\frac{\mu_\circ^4}{V"_{\!\!\circ}}
&\approx &
\dis 10^{-16}\, \left(\frac{\mu_\circ(\hbox{TeV})}{\Lambda_{\rm inflation }/(10^{16}\ \hbox{GeV})}\right)^{\!4}\,.
\ea
\ee

\vspace{2mm}
\noindent
$\mu_\circ\approx 3$  TeV, for example, would then lead  to an E\"otv\"os parameter  $\,\delta\,\approx\,10^{-14}$ to $10^{-16}$, \,for  $\Lambda_{\rm inflation }\approx $ (1 to 3)$\,\times\, 10^{16}$ GeV.
\vspace{2mm}

The very weak strength of the gravitational interaction for individual particles is usually associated with the very large Planck scale $\,\simeq\, 10^{19}$ GeV. The even tinier strength of a new interaction, smaller than gravity or electromagnetism by  
 $\,\simge 10^{12}$ or
 $\,\simge 10^{48}$ as required by Equivalence Principle tests, may be associated under  a 
\vspace{-1mm}
 \be
\hbox {\it very high energy \ \,$\longleftrightarrow$ \ very small coupling \ connection\,} 
 \ee
 
 \vspace{1mm}
 \noindent
 with a new energy scale $\sqrt \xi$\,, through a very large $\xi D$ term within supersymmetry.

 \pagebreak
 \vspace{2mm}
 
 This scale  $\sqrt \xi$  must be $\simge 10^{12}$ larger  than the $\sim$ few TeV usually associated with supersymmetry breaking, thus $\simge \,10^{15}$\,--\,$10^{16}$ GeV. Its associated vacuum energy density, huge, may be at the origin of the very rapid inflation of the early Universe. Such a new interaction, although extremely weak, could still be accessible to improved tests of the Equivalence Principle.

%%\pagebreak

\ghost{
\bc
* \hspace{1mm}  * \hspace{1mm}  *
\ec
}
 
 \section{Conclusion}
 
Altogether the first results of the {\it MICROSCOPE\,} experiment, which provide the best test of the Equivalence Principle, 
lead to improved (2$\sigma$)  limits on the strength of a new long-range force as compared to gravity, ranging from
$|\bar\alpha| < 3.2\times 10^{-11}$ for a coupling to $B\,$ to (2.3 or 2.2)\,$\times \,10^{-13}$  for a coupling to $L$ or $B-L$, and $6.7 \times 10^{-13}$ for a coupling to $B+L$ (with, in the spin-1 case, a coupling to $B-L$ favored by grand unification). The corresponding limits on $\epsilon$, parametrizing the strength of the couplings as compared to $e$, range from $5 \times 10^{-24}$ for $\epsilon_B$ to less than $10^{-24}$ in the other cases. Such an extremely small coupling may be associated with a very large energy scale, corresponding to a huge energy density that may be responsible for the inflation of the early Universe.

\vspace{6mm}
\noindent
Acknowledgements:

\vspace{2mm}
I thank the referees for very useful comments and suggestions.

\bibliography{References}

\end{document}